\documentclass[
  aps,
  prb,
  reprint,
  superscriptaddress,
  nofootinbib,
  longbibliography
]{revtex4-2}
\usepackage{amsmath,amssymb}
\usepackage{graphicx}
\DeclareGraphicsExtensions{.pdf,.png,.jpg,.jpeg}
\usepackage{placeins}
\usepackage{hyperref}
\hypersetup{hidelinks}

\allowdisplaybreaks[2]

\newcommand{\dd}{\mathrm{d}}
\newcommand{\ii}{\mathrm{i}}
\newcommand{\ee}{\mathrm{e}}
\newcommand{\br}{\mathbf r}
\newcommand{\bR}{\mathbf R}
\newcommand{\bT}{\mathbf T}
\newcommand{\bk}{\mathbf k}
\newcommand{\bq}{\mathbf q}
\newcommand{\bQ}{\mathbf Q}
\newcommand{\bG}{\mathbf G}
\newcommand{\ba}{\mathbf a}
\newcommand{\bb}{\mathbf b}
\newcommand{\btau}{\boldsymbol\tau}
\newcommand{\brho}{\boldsymbol\rho}
\newcommand{\dk}{\Delta\mathbf k}
\newcommand{\dkap}{\Delta\boldsymbol\kappa}
\newcommand{\PAW}{\mathrm{PAW}}
\newcommand{\ps}{\mathrm{ps}}
\newcommand{\cell}{\mathrm{cell}}
\newcommand{\BvK}{\mathrm{BvK}}

\graphicspath{{./}}

\begin{document}

\title{Beyond Local Berry Geometry: A First-Principles Finite-Momentum Theory of Electronic Position}

\author{M. S. Si}
\email{sims@lzu.edu.cn}
\affiliation{School of Materials and Energy, Lanzhou University, Lanzhou 730000, China}

\author{Y. Q. Liu}
\affiliation{School of Materials and Energy, Lanzhou University, Lanzhou 730000, China}

\author{G. P. Zhang}
\email{guo-ping.zhang@outlook.com}
\affiliation{Department of Physics, Indiana State University, Terre Haute, Indiana 47809, USA}

\date{\today}

\begin{abstract}
Electronic position controls how a crystal polarizes and responds to an
external field. In crystals, it is usually described through local changes
of electronic states in momentum space. This Berry framework has reshaped
modern solid-state physics, but strong fields drive electrons across a
finite momentum range, where coherence between different momenta becomes
part of the response. Here we establish a first-principles theory of
electronic position at finite momentum that retains this missing
information. We show that unequal-momentum coherence can cancel under
spatial averaging and still produce polarization, forming a coherence
dipole. We obtain the matrix directly from material wave functions, without
model bands or fitted transition elements. In Si, the finite-momentum
geometry sets a material momentum scale. Comparing this scale with the
momentum change driven by the field predicts when finite-momentum physics
becomes active. Crossing the scale strongly reorganizes the fifth and
higher harmonics, showing that momentum-space geometry, rather than emitted
photon energy alone, controls the nonlinear response. HHG is the first
demonstration, but the theory applies whenever driven electrons explore a
finite momentum range. It therefore extends quantum geometry beyond the
local Berry limit and provides a general basis for predicting field-driven
phenomena in real materials.
\end{abstract}

\maketitle

\section{Introduction}

Electronic position determines how a solid polarizes, carries current, and
emits light. In a crystal, position cannot be treated in the same simple way
as in a finite system. Berry geometry solved the local part of this problem
by linking observable response to changes of Bloch states at nearby crystal
momenta
\cite{aversa1995,sipe2000,nunes2001,souza2002,souza2004,
ghimire2011,vampa2015}. This local picture now supports modern theories of
polarization, charge motion, and nonlinear optical response. Strong fields,
however, move electrons across a finite momentum range and make coherence
between different momenta part of the response
\cite{vampa2015,vampa2014,freeman2022,reislohner2022}. Electronic position
at finite momentum is therefore the missing link between local Berry
geometry and the dynamics of driven materials.

Here we establish a first-principles theory that keeps electronic position
between different crystal momenta. Its finite-momentum matrix contains the
local Berry connection but also retains coherence lost from the local
description. These matrix elements are not small numerical corrections;
they describe a part of the electronic response that the local theory does
not contain. We show that this coherence can cancel under spatial averaging
while retaining a finite position-weighted moment, forming a coherence
dipole. The theory also gives a material momentum scale that can be compared
directly with the motion driven by an external field. This comparison shows
when finite-momentum physics becomes active and explains why photon energy
alone does not set the response. Because the theory starts from
first-principles wave functions, it connects this physics directly to real
materials and measurable dynamics. HHG is the first demonstration; the same
finite-momentum information can control a broad range of field-driven
electronic responses.

The challenge follows directly from the length-gauge coupling of a
spatially uniform electric field,
\begin{equation}
    \hat H_I(t)=e\,\mathbf E(t)\cdot\hat{\mathbf R},
\end{equation}
where \(e>0\) is the elementary charge. This length-gauge form makes
polarization and field-driven motion explicit, but \(R_\alpha\) is not
lattice periodic. In the continuum Bloch representation, position appears
as a derivative of the momentum delta distribution together with a Berry
connection \cite{blount1962,si2025position}. Berry-phase, overlap-link,
Wannier, and covariant-derivative methods use this local structure to obtain
well-defined bulk observables
\cite{nunes2001,souza2002,kingsmith1993,resta1998,marzari1997,
valenca2019,ventura2017,ditler2023,esteve2023}.
In the local limit, finite-separation overlaps recover Berry geometry; at
finite \(Q\), their amplitude and phase retain additional information.

On a finite Bloch manifold, crystal momentum is represented by a discrete
set and the electronic states span a finite band subspace. For a
BvK-compatible mesh
\(\mathbf N=(N_1,N_2,N_3)\), the corresponding operator is
\begin{equation}
    \hat R_\alpha^{(\mathbf N)}
    =\hat{\mathcal P}_{\mathbf N}\hat R_\alpha
     \hat{\mathcal P}_{\mathbf N}.
\end{equation}
Its matrix is not generally diagonal in crystal momentum. The
unequal-momentum blocks collectively represent intercell displacement and
the momentum derivative that generates Bloch acceleration. They enter one
position operator and one coherent polarization. Consistent band frames
and operator projections preserve this gauge-covariant structure
\cite{ventura2017,taghizadeh2017,passos2018,yue2020,yue2022,
puente2024,an2026}.

\section{Finite-momentum electronic position}

We fix a finite BvK domain \(V_{\BvK}\) and a single coordinate branch
within it. Let \(\mathbb K_{\mathbf N}\) denote a uniformly shifted,
BvK-compatible \(k\)-point mesh,
\begin{equation}
\bk_{\mathbf j}=\sum_{\mu=1}^{3}
\frac{j_\mu+\theta_\mu}{N_\mu}\bb_\mu,
\qquad
j_\mu=0,\ldots,N_\mu-1,
\label{eq:shifted_bvk_mesh}
\end{equation}
where the common twist \(\boldsymbol\theta\) allows a shifted mesh and
cancels from every momentum difference. We use BvK-normalized Bloch states
\begin{equation}
\psi_{n\bk}(\bR)=\frac{1}{\sqrt{N_R}}
\ee^{\ii\bk\cdot\bR}u_{n\bk}(\bR),
\qquad
\langle u_{n\bk}|u_{m\bk}\rangle_{\cell}=\delta_{nm},
\label{eq:bvk_bloch_normalization}
\end{equation}
with \(u_{n\bk}(\bR+\bT)=u_{n\bk}(\bR)\). The projected operator is
\begin{equation}
    \hat R_{\alpha}^{(\mathbf N)}=\hat{\mathcal P}_{\mathbf N}\hat R_\alpha\hat{\mathcal P}_{\mathbf N},
    \qquad
    \hat{\mathcal P}_{\mathbf N}
    =\sum_{\bk\in\mathbb K_{\mathbf N}}\sum_{n=1}^{N_b}
    |\psi_{n\bk}\rangle\langle\psi_{n\bk}|,
\label{eq:finite_projected_operator}
\end{equation}
where \(N_b\) is the number of retained bands. For states in this subspace,
\begin{align}
    \mathcal R^\alpha_{nm}(\bk,\bq)
    & =\langle\psi_{n\bk}|\hat R_{\alpha}^{(\mathbf N)}|\psi_{m\bq}\rangle \nonumber\\
    & =\int_{V_{\BvK}}\!\dd^3R\,
      \psi^*_{n\bk}(\bR)R_\alpha\psi_{m\bq}(\bR).
\label{eq:finite_matrix_definition}
\end{align}
This definition does not make $\hat R_\alpha$ lattice periodic. The matrix depends 
on the chosen finite coordinate branch, and no condition $n=m$ or $\bk=\bq$ is 
imposed before integration.

\subsection{Finite domain and notation}

We use a parallelepipedal coordinate branch with vertex \(\bR_0=0\). 
A different origin or coordinate-branch convention requires the corresponding 
coordinate shift and basis transformation; all coordinates below refer 
to this branch.

The primitive-cell lattice vectors are denoted by \(\ba_\mu\), with \(\mu=1,2,3\). Indices \(\mu,\nu,d\in\{1,2,3\}\) 
label primitive-lattice directions, \(A\) labels atoms, \(i,j\) label PAW projector channels,
 \(n,m\) label the original bands, and \(a,b\) label bands in the aligned frame. 
 The Cartesian component is denoted by \(\alpha\in\{x,y,z\}\).
The reciprocal vectors \(\bb_\mu\) satisfy
\begin{equation}
    \ba_\mu\cdot\bb_\nu=2\pi\delta_{\mu\nu}.
\end{equation}
A point inside the primitive cell is written as
\begin{equation}
\br=\sum_{\mu=1}^{3}s_\mu\ba_\mu,
\qquad 0\le s_\mu<1.
\label{eq:position_basis}
\end{equation}
Here \(s_\mu\) is the dimensionless fractional coordinate 
along \(\ba_\mu\), so that \(\br\cdot\bb_\mu=2\pi s_\mu\).
When a single direction is selected, \(s_d\) denotes the 
corresponding fractional coordinate, not a Cartesian length. 
A finite BvK supercell contains \(N_R=N_1N_2N_3\) primitive cells, 
and the translation of a cell is
\begin{equation}
\bT=\sum_{\mu=1}^{3}h_\mu\ba_\mu,
\qquad h_\mu=0,\ldots,N_\mu-1.
\label{eq:translation_vector}
\end{equation}
The full coordinate is then
\begin{equation}
    \bR=\bT+\br.
\label{eq:coordinate_partition}
\end{equation}
Figure~\ref{fig:si}(a) illustrates the partition: $\bT$ locates a 
primitive cell within the finite domain, 
$\br$ lies inside that cell, and $\bR$ is their sum. 
The derivation uses primitive vectors and a primitive-cell partition, 
independent of the conventional-cell view used to display diamond Si.
For the Cartesian component \(\alpha\),
\begin{equation}
    R_\alpha=T_\alpha+r_\alpha,
\qquad
    r_\alpha=\sum_{\mu=1}^{3}a_{\alpha\mu}s_\mu,
\label{eq:cartesian_component}
\end{equation}
where \(a_{\alpha\mu}\) is the \(\alpha\)-component of \(\ba_\mu\).

We distinguish the Cartesian momentum difference \(\dk=\bq-\bk\) 
from its reduced reciprocal-lattice coordinates \(\dkap\), defined by
\begin{equation}
    \bq-\bk=\sum_{\mu=1}^{3}\Delta\kappa_\mu\bb_\mu.
\label{eq:reduced_delta_k}
\end{equation}
We also write \(\bk=\sum_\mu\kappa_\mu\bb_\mu\) and \(\bq=\sum_\mu\eta_\mu\bb_\mu\), 
so that \(\Delta\kappa_\mu=\eta_\mu-\kappa_\mu\), up to the
reciprocal-lattice representative chosen at a Brillouin-zone boundary. 
Each reciprocal-torus point is represented once; throughout, equal
\(k\) means equality modulo a reciprocal-lattice vector.
Accordingly, \(\partial/\partial\kappa_d\) denotes
differentiation along \(\bb_d\), with the other reduced coordinates held fixed.
Likewise, a reciprocal lattice vector is written as
\begin{equation}
    \bG=\sum_{\mu=1}^{3}g_\mu\bb_\mu,
\qquad g_\mu\in\mathbb Z.
\end{equation}
With these definitions,
\begin{equation}
    \ee^{\ii(\bq-\bk)\cdot\bT}
    =
    \ee^{\ii2\pi\sum_\mu\Delta\kappa_\mu h_\mu}.
\end{equation}
The pseudo cell-periodic state is expanded as
\begin{equation}
    \tilde u_{n\bk}(\br)=\frac{1}{\sqrt{\Omega_{\cell}}}
    \sum_{\bG}C_{n\bk}(\bG)\ee^{\ii\bG\cdot\br}.
\label{eq:plane_wave_expansion}
\end{equation}
The coefficients \(C_{n\bk}(\bG)\) are obtained from the plane-wave
 pseudo-wavefunction. Throughout, \(\tilde u_{n\bk}\) denotes 
 the pseudo cell-periodic state, whereas \(u_{n\bk}\) denotes 
 the corresponding all-electron valence state reconstructed within the
 frozen-core PAW approximation.
  We use the cell inner product \(\langle f|g\rangle_{\cell}=\int_{\cell}\dd^3r\,f^*(\br)g(\br)\) 
  and normalize the all-electron states as \(\langle u_{n\bk}|u_{m\bk}\rangle_{\cell}=\delta_{nm}\). 
  With the normalized plane-wave convention in Eq.~\eqref{eq:plane_wave_expansion},
  \(\sum_{\bG}C^*_{n\bk}(\bG)C_{m\bk}(\bG)=\langle\tilde u_{n\bk}|\tilde u_{m\bk}\rangle_{\cell}\),
  where \(\Omega_{\cell}\) is the primitive-cell volume.

\subsection{Direct finite-domain decomposition}

For retained states, the projected and unprojected matrix elements 
coincide within the chosen finite BvK domain:
\begin{equation}
\begin{aligned}
    \mathcal R^\alpha_{nm}(\bk,\bq)
    &\equiv
    \langle \psi_{n\bk}|\hat R_\alpha^{(\mathbf N)}|\psi_{m\bq}\rangle \\
    &=
    \langle \psi_{n\bk}|\hat R_\alpha|\psi_{m\bq}\rangle_{\BvK}.
\end{aligned}
\label{eq:BvK_int}
\end{equation}
The second equality holds because both states belong to the 
retained subspace. Substituting the Bloch functions and 
using \(\bR=\bT+\br\), one obtains
\begin{align}
    \mathcal R^\alpha_{nm}(\bk,\bq)
    ={}&
    \frac{1}{N_R}
    \sum_{\bT}
    \ee^{\ii(\bq-\bk)\cdot\bT} \nonumber\\
    &\times\left[
    T_\alpha S_{nm}(\bk,\bq)
    +I^\alpha_{nm}(\bk,\bq)
    \right],
\label{eq:BvK_po}
\end{align}
where
\begin{equation}
S_{nm}(\bk,\bq)
=
\left\langle u_{n\bk}\middle|
\ee^{\ii(\bq-\bk)\cdot\br}
\middle|u_{m\bq}
\right\rangle_{\cell},
\label{eq:primitive_overlap_general}
\end{equation}
and
\begin{equation}
I^\alpha_{nm}(\bk,\bq)
=
\left\langle
u_{n\bk}\middle|
\ee^{\ii(\bq-\bk)\cdot\br} r_\alpha
\middle|u_{m\bq}
\right\rangle_{\cell}.
\label{eq:primitive_position_general}
\end{equation}
The complex conjugation of the first state is 
implicit in the inner product. 
The primitive-cell integral \(I^\alpha_{nm}(\bk,\bq)\) is defined for a
general pair of mesh \(k\) points. 
The corresponding 
pseudo-wavefunction quantities are obtained
 by replacing \(u\) with \(\tilde u\):
\begin{equation}
S^{\ps}_{nm}(\bk,\bq)
=
\left\langle \tilde u_{n\bk}\middle|
\ee^{\ii(\bq-\bk)\cdot\br}
\middle|\tilde u_{m\bq}
\right\rangle_{\cell},
\label{eq:primitive_overlap_ps}
\end{equation}
and
\begin{equation}
I^{\alpha,\ps}_{nm}(\bk,\bq)
=
\left\langle
\tilde u_{n\bk}\middle|
\ee^{\ii(\bq-\bk)\cdot\br}r_\alpha
\middle|\tilde u_{m\bq}
\right\rangle_{\cell}.
\label{eq:primitive_position_ps}
\end{equation}

The ordinary translational phase average and the 
position-weighted translational sum are defined as
\begin{equation}
S_T(\dkap)
=
\frac{1}{N_R}
\sum_{\bT}
\ee^{\ii2\pi\sum_\mu\Delta\kappa_\mu h_\mu},
\label{eq:ST_def}
\end{equation}
\begin{equation}
\mathbf R_{\mathrm{sum}}(\dkap)
=
\frac{1}{N_R}
\sum_{\bT}
\bT\,
\ee^{\ii2\pi\sum_\mu\Delta\kappa_\mu h_\mu}.
\label{eq:Rsum_def}
\end{equation}
Writing \(\bQ=\bq-\bk\), the two lattice sums obey
\begin{align}
\mathbf R_{\mathrm{sum}}(\bQ)
&=\frac{1}{\ii}\boldsymbol\nabla_{\bQ}S_T(\bQ),\nonumber\\
\mathbf d_{\mathrm{coh}}^{nm}(\bk,\bq)
&=\mathbf R_{\mathrm{sum}}(\bQ)S_{nm}(\bk,\bq).
\label{eq:coherence_dipole}
\end{align}
Equation~\eqref{eq:coherence_dipole} exposes the central physical content:
an unequal-momentum coherence can have a vanishing spatial average while
retaining a finite first moment. It is dark to the zeroth spatial moment but
bright to electronic position, forming a coherence dipole that contributes
directly to polarization.
Equation~\eqref{eq:BvK_po} may therefore be expressed as
\begin{equation}
    \mathcal R^\alpha_{nm}(\bk,\bq)
    =
    R^\alpha_{\mathrm{sum}}(\dkap)S_{nm}(\bk,\bq)
    +
    S_T(\dkap)I^\alpha_{nm}(\bk,\bq).
\label{eq:direct_matrix_general}
\end{equation}
Equation~\eqref{eq:direct_matrix_general} is the central finite-mesh
relation. Its first term is the intercell first moment: a
position-weighted translational sum multiplying a primitive-cell overlap.
The second term is the intracell coordinate matrix multiplied by the
ordinary translational phase average. Their relative weight is fixed by
the lattice sums rather than by an imposed \(k\)-selection rule.

\subsection{Equal-\(k\) pseudo-wavefunction block}

For \(\bq=\bk\), one has \(\dkap=0\),
\begin{equation}
S_T(0)=1,
\qquad
\mathbf R_{\mathrm{sum}}(0)=\frac{1}{N_R}\sum_{\bT}\bT\equiv \mathbf T_c.
\end{equation}
For a regular BvK-compatible mesh,
\begin{equation}
\mathbf T_c=\frac{N_1-1}{2}\ba_1+\frac{N_2-1}{2}\ba_2+\frac{N_3-1}{2}\ba_3.
\label{eq:Tc_def}
\end{equation}
Thus, \(\mathbf T_c\) is the average translation vector, or equivalently 
the center of the set of primitive-cell origins. It changes when 
the coordinate branch is changed. The equal-\(k\) 
pseudo-wavefunction contribution is
\begin{equation}
\mathcal R^{\alpha,\ps}_{nm}(\bk,\bk)
=
I^{\alpha,\ps}_{nm}(\bk,\bk)+T_c^\alpha S^{\ps}_{nm}(\bk,\bk),
\label{eq:equal_k_ps_total}
\end{equation}
where
\begin{equation}
S^{\ps}_{nm}(\bk,\bk)=\sum_{\bG}C^*_{n\bk}(\bG)C_{m\bk}(\bG).
\end{equation}
The primitive-cell position integral is
\begin{equation}
I^{\alpha,\ps}_{nm}(\bk,\bk)
=
\sum_{\bG,\bG'}C^*_{n\bk}(\bG)C_{m\bk}(\bG')K^\alpha_{\mathrm{cell}}(\bG'-\bG),
\end{equation}
with
\begin{equation}
K^\alpha_{\mathrm{cell}}(\Delta\bG)
=
\frac{1}{\Omega_{\cell}}
\left\langle
\ee^{\ii\bG\cdot\br}
\middle|r_\alpha\middle|
\ee^{\ii\bG'\cdot\br}
\right\rangle_{\cell}.
\end{equation}
Using Eq.~\eqref{eq:position_basis},
\begin{align}
K^\alpha_{\mathrm{cell}}(\Delta\bG)
={}&\sum_{\mu=1}^{3}a_{\alpha\mu}
\int_0^1\dd s_1\dd s_2\dd s_3\,s_\mu \nonumber\\
&\times\ee^{\ii2\pi(\Delta g_1s_1+\Delta g_2s_2+\Delta g_3s_3)},
\label{eq:Kcell_integral}
\end{align}
where \(\Delta g_\mu=g'_\mu-g_\mu\). Since the integral is separable, 
the contribution from direction \(\mu\) is nonzero only when the other 
two reciprocal-lattice differences satisfy
\begin{equation}
\Delta g_\nu=0,\qquad \nu\ne\mu.
\end{equation}
The remaining one-dimensional integral is
\begin{equation}
J(\lambda)=\int_0^1 s\ee^{\ii2\pi\lambda s}\dd s
=
\begin{cases}
1/2,& \lambda=0,\\[4pt]
-\ii/(2\pi\lambda),& \lambda\ne0,\quad \lambda\in\mathbb Z.
\end{cases}
\label{eq:J_integral}
\end{equation}

\subsection{PAW-reconstructed equal-\(k\) block}

For a general operator \(\hat{\mathcal O}\), the PAW matrix element is \cite{blochl1994,kresse1999}
\begin{equation}
\langle \psi_n|\hat{\mathcal O}|\psi_m\rangle
=
\langle \tilde\psi_n|\hat{\mathcal O}|\tilde\psi_m\rangle
+
\sum_A\sum_{ij}
\langle \tilde\psi_n|\tilde p_i^A\rangle
\Delta\mathcal O^A_{ij}
\langle \tilde p_j^A|\tilde\psi_m\rangle,
\label{eq:paw_general}
\end{equation}
where
\begin{equation}
\Delta\mathcal O^A_{ij}
=
\langle \phi_i^A|\hat{\mathcal O}|\phi_j^A\rangle
-
\langle \tilde\phi_i^A|\hat{\mathcal O}|\tilde\phi_j^A\rangle.
\end{equation}
For atom \(A\) in the cell translated by \(\bT\), its position is \(\bT+\btau_A\).  
The local coordinate in the augmentation sphere is
\begin{equation}
\brho=\bR-(\bT+\btau_A),
\qquad
\bR=\bT+\btau_A+\brho.
\end{equation}
We use Bloch-summed PAW projectors
\begin{equation}
|\tilde p_{i\bk}^{A}\rangle
=\frac{1}{\sqrt{N_R}}\sum_{\bT}
\ee^{\ii\bk\cdot\bT}|\tilde p_i^{A,\bT}\rangle ,
\label{eq:bloch_summed_paw_projector}
\end{equation}
and retain the atomic phase associated with \(\btau_A\) in the cell
coefficient
\(P_{i,n\bk,A}=\langle\tilde p_{i\bk}^{A}|
\tilde\psi_{n\bk}\rangle\).
For equal-\(k\) matrix elements, the averaged translation
 is \(\mathbf T_c\).  Thus the PAW position correction on atom \(A\) is
\begin{equation}
\Delta\mathbf R^A_{ij}
=(\btau_A+\mathbf T_c)\Delta S^A_{ij}+\Delta\mathbf d^A_{ij},
\label{eq:equal_k_paw_onsite}
\end{equation}
where
\begin{equation}
\Delta S^A_{ij}
=
\langle\phi_i^A|\phi_j^A\rangle
-
\langle\tilde\phi_i^A|\tilde\phi_j^A\rangle,
\end{equation}
\begin{equation}
\Delta\mathbf d^A_{ij}
=
\langle\phi_i^A|\brho|\phi_j^A\rangle
-
\langle\tilde\phi_i^A|\brho|\tilde\phi_j^A\rangle.
\end{equation}
The equal-\(k\) PAW matrix element is then
\begin{align}
\mathcal R^{\alpha,\PAW}_{nm}(\bk,\bk)
={}&\mathcal R^{\alpha,\ps}_{nm}(\bk,\bk) \nonumber\\
&+\sum_A\sum_{ij}
P^*_{i,n\bk,A}\,\Delta R^{A,\alpha}_{ij}\,P_{j,m\bk,A}.
\label{eq:equal_k_paw_total}
\end{align}
Here, the coefficient \(P_{i,n\bk,A}\) follows the convention in
Eq.~\eqref{eq:bloch_summed_paw_projector}.

For two all-electron Bloch states at equal \(k\), 
define the cell-periodic coherence density
\begin{equation}
\mathcal C_{nm\bk}(\br)
=
 u^*_{n\bk}(\br)u_{m\bk}(\br).
\label{eq:cell_coherence_density}
\end{equation}
The common Bloch phase cancels, so \(\mathcal C_{nm\bk}(\br)\) repeats with 
the same phase and spatial form in every primitive cell. All-electron orthonormality gives
\begin{equation}
\int_{\cell}\mathcal C_{nm\bk}(\br)\,\dd^3r=\delta_{nm}.
\end{equation}
After subtracting the average-translation contribution, we define
\begin{equation}
\mathcal K^\alpha_{nm}(\bk)
\equiv
\mathcal R^{\alpha,\PAW}_{nm}(\bk,\bk)
-T_c^\alpha\delta_{nm}
=
\int_{\cell}r_\alpha\mathcal C_{nm\bk}(\br)\,\dd^3r.
\label{eq:coherence_displacement_kernel}
\end{equation}
The last equality is understood at the all-electron PAW level,
 where the pseudo term and the on-site reconstruction in Eq.~\eqref{eq:equal_k_paw_total} form one matrix element.
 We refer to \(\mathcal K^\alpha_{nm}(\bk)\) as the \emph{equal-\(k\) intracell displacement}; 
 for \(n\ne m\), it is the \emph{intracell coherence-displacement}.

Its real-space meaning follows by contracting with the density matrix. 
For \(n\ne m\), an equal-\(k\) coherence produces the following 
real, cell-periodic density redistribution per spin
\begin{equation}
\delta n_{nm\bk}(\br,t)
=
\frac{2}{N_R}\,\operatorname{Re}\!\left[
\rho_{mn}(\bk,\bk;t)\mathcal C_{nm\bk}(\br)
\right].
\label{eq:coherence_charge_redistribution}
\end{equation}
Its cell integral vanishes by orthogonality, whereas its 
position-weighted integral---the first spatial moment---is 
generally finite. The corresponding pair contribution 
to the electronic charge displacement is
\begin{equation}
\delta p^\alpha_{nm\bk}(t)
=
-\frac{2g_s e}{N_R}\,\operatorname{Re}\!\left[
\rho_{mn}(\bk,\bk;t)\mathcal K^\alpha_{nm}(\bk)
\right].
\label{eq:coherence_displacement_moment}
\end{equation}
Here the expression refers to the contribution per primitive cell from one
unordered band pair, for example \(n<m\): the factor 2 combines the
Hermitian-conjugate pair, \(1/N_R\) is the uniform mesh weight, and \(g_s\)
accounts for spin degeneracy. The pseudo term describes 
the smooth valence and interstitial contributions, whereas PAW 
augmentation restores the on-site all-electron part.

\subsection{Unequal-\(k\) selection rule}

For a regular BvK-compatible mesh,
\begin{equation}
\Delta\kappa_\mu=\frac{c_\mu}{N_\mu}+z_\mu,
\qquad c_\mu,z_\mu\in\mathbb Z.
\end{equation}
Define
\begin{equation}
\begin{aligned}
\mathcal A_\mu(\xi)
&=\frac{1}{N_\mu}\sum_{h_\mu=0}^{N_\mu-1}\ee^{\ii2\pi\xi h_\mu},\\
\mathcal B_\mu(\xi)
&=\frac{1}{N_\mu}\sum_{h_\mu=0}^{N_\mu-1}h_\mu\ee^{\ii2\pi\xi h_\mu}.
\end{aligned}
\end{equation}
On this mesh, \(\mathcal A_\mu(\xi)=1\) for integer \(\xi\) and vanishes for \(\xi=c_\mu/N_\mu+z_\mu\) with \(c_\mu\not\equiv0\pmod{N_\mu}\). Hence
\begin{equation}
S_T(\dkap)=\prod_{\mu=1}^{3}\mathcal A_\mu(\Delta\kappa_\mu).
\label{eq:ST_product}
\end{equation}
Thus, on a regular BvK-compatible mesh, \(S_T\) vanishes whenever at 
least one component \(\Delta\kappa_\mu\) is noninteger.

The position-weighted translational sum may be expressed as
\begin{equation}
\mathbf R_{\mathrm{sum}}(\dkap)
=
\sum_{d=1}^{3}\ba_d\,
\mathcal B_d(\Delta\kappa_d)
\prod_{\mu\ne d}\mathcal A_\mu(\Delta\kappa_\mu).
\label{eq:Rsum_factorized}
\end{equation}
For a noninteger BvK-compatible component \(\xi\),
\begin{equation}
\mathcal B_\mu(\xi)=-\frac{1}{1-\ee^{\ii2\pi\xi}},
\label{eq:calB_noninteger}
\end{equation}
whereas \(\mathcal B_\mu(\xi)=(N_\mu-1)/2\) when \(\xi\in\mathbb Z\). 
Equation~\eqref{eq:Rsum_factorized} then shows that \(\mathbf R_{\mathrm{sum}}\) 
is nonzero for unequal-\(k\) pairs only when exactly one reduced component 
is noninteger and the other two are integers. If the noninteger component lies along direction \(d\),
\begin{equation}
\mathbf R_{\mathrm{sum}}(\dkap)
=
\ba_d\left[-\frac{1}{1-\ee^{\ii2\pi\Delta\kappa_d}}\right],
\label{eq:Rsum_allowed_neq}
\end{equation}
provided the two transverse components are integer-compatible. 
If two or three reduced components are noninteger, \(\mathbf R_{\mathrm{sum}}\) vanishes.

\subsection{Allowed unequal-\(k\) pseudo-wavefunction block}

For an allowed unequal-\(k\) pair whose noninteger component lies along direction \(d\),
 the ordinary phase average vanishes and Eq.~\eqref{eq:direct_matrix_general} reduces to
\begin{equation}
\mathcal R^{\alpha,\ps}_{nm}(\bk,\bq)
=
R^\alpha_{\mathrm{sum}}(\dkap)S^{\ps}_{nm}(\bk,\bq).
\label{eq:neq_ps_basic}
\end{equation}
The primitive-cell overlap in the plane-wave basis is
\begin{equation}
S^{\ps}_{nm}(\bk,\bq)
=
\sum_{\bG,\bG'}{}'
C^*_{n\bk}(\bG)C_{m\bq}(\bG')
L(g'_d-g_d+\Delta\kappa_d),
\label{eq:neq_overlap}
\end{equation}
where the prime restricts the reciprocal-vector sum to terms satisfying
\begin{equation}
 g'_\mu-g_\mu+\Delta\kappa_\mu=0,
\qquad \mu\ne d.
\label{eq:transverse_compensation}
\end{equation}
The one-dimensional integral in the noninteger direction is
\begin{equation}
L(\delta)=\int_0^1\ee^{\ii2\pi\delta s}\dd s
=
\begin{cases}
1,& \delta=0,\\[4pt]
\dfrac{\ee^{\ii2\pi\delta}-1}{\ii2\pi\delta},& \delta\ne0.
\end{cases}
\label{eq:L_integral}
\end{equation}
Combining Eqs.~\eqref{eq:Rsum_allowed_neq} and \eqref{eq:neq_overlap}, the Cartesian component of the unequal-\(k\) pseudo-wavefunction matrix element is
\begin{align}
\mathcal R^{\alpha,\ps}_{nm}(\bk,\bq)
={}&a_{\alpha d}
\left[-\frac{1}{1-\ee^{\ii2\pi\Delta\kappa_d}}\right] \nonumber\\
&\times\sum_{\bG,\bG'}{}'
C^*_{n\bk}(\bG)C_{m\bq}(\bG') \nonumber\\
&\times L(g'_d-g_d+\Delta\kappa_d).
\label{eq:neq_ps_final}
\end{align}
Since \(\lambda=g'_d-g_d\in\mathbb Z\), the translational factor and the
primitive-cell integral satisfy, for \(\delta\notin\mathbb Z\), the identity
\begin{equation}
-\frac{1}{1-\ee^{\ii2\pi\delta}}
\frac{\ee^{\ii2\pi(\lambda+\delta)}-1}{\ii2\pi(\lambda+\delta)}
=
\frac{1}{\ii2\pi(\lambda+\delta)}.
\label{eq:finite_kernel_identity}
\end{equation}
The resulting finite unequal-\(k\) kernel is
\begin{equation}
F_\lambda(\delta)=\frac{1}{\ii2\pi(\lambda+\delta)}.
\label{eq:finite_kernel}
\end{equation}
For \(\lambda\ne0\), \(F_\lambda(\delta)\to -\ii/(2\pi\lambda)=J(\lambda)\) as \(\delta\to0\), so the regular Fourier components 
join continuously to the equal-\(k\) cell-coordinate kernel. The exceptional term \(F_0(\delta)=1/(\ii2\pi\delta)\) 
controls the mesh scaling of near-diagonal unequal-\(k\) blocks, discussed in Sec.~\ref{sec:consistency}.

\subsection{PAW-reconstructed unequal-\(k\) block}

With the projector convention of
Eq.~\eqref{eq:bloch_summed_paw_projector}, the augmentation part for a
general pair can be written as
\begin{align}
\Delta\boldsymbol{\mathcal R}^{A}_{nm}(\bk,\bq)
={}&\sum_{ij}P^*_{i,n\bk,A}
\big\{\mathbf R_{\mathrm{sum}}(\dkap)\Delta S^A_{ij}
\nonumber\\
&\quad+S_T(\dkap)
[\btau_A\Delta S^A_{ij}+\Delta\mathbf d^A_{ij}]
\big\}P_{j,m\bq,A}.
\label{eq:general_paw_position_augmentation}
\end{align}
For allowed unequal-\(k\) matrix elements, the local-coordinate and
atomic-position terms are proportional to \(S_T(\dkap)\) and therefore
vanish. The surviving PAW term is the position-weighted translational sum
multiplied by the on-site overlap correction,
\begin{equation}
\Delta\mathbf R^{A,\mathrm{neq}}_{ij}
=
\mathbf R_{\mathrm{sum}}(\dkap)\Delta S^A_{ij}.
\label{eq:neq_paw_onsite}
\end{equation}
Neither the atomic-position term \(\btau_A\) nor the local-coordinate term involving \(\brho\) appears in Eq.~\eqref{eq:neq_paw_onsite}, 
because both belong to the phase-average term removed by \(S_T=0\). The
complete frozen-core PAW valence matrix element is
\begin{align}
\mathcal R^{\alpha,\PAW}_{nm}(\bk,\bq)
={}&\mathcal R^{\alpha,\ps}_{nm}(\bk,\bq) \nonumber\\
&+\sum_A\sum_{ij}P^*_{i,n\bk,A} \nonumber\\
&\quad\times R^\alpha_{\mathrm{sum}}(\dkap)\Delta S^A_{ij}P_{j,m\bq,A}.
\label{eq:neq_paw_total}
\end{align}
Here
\begin{equation}
P_{i,n\bk,A}=\langle \tilde p_{i\bk}^A|\tilde\psi_{n\bk}\rangle,
\qquad
P_{j,m\bq,A}=\langle \tilde p_{j\bq}^A|\tilde\psi_{m\bq}\rangle
\label{eq:paw_projector_coefficients}
\end{equation}
are coefficients of the Bloch-summed PAW projectors. 
The pseudo and augmentation terms use the same phase convention. Here the
atomic Bloch phase
\(\exp[\ii(\bq-\bk)\cdot\btau_A]\) is retained in the cell coefficients;
consequently, no additional atomic phase multiplies
Eq.~\eqref{eq:neq_paw_total}.

To expose the real-space content of the unequal-\(k\) block, define the transition density
\begin{equation}
\varrho_{n\bk,m\bq}(\bR)=\psi^*_{n\bk}(\bR)\psi_{m\bq}(\bR).
\end{equation}
Under a lattice translation it obeys
\begin{equation}
\varrho_{n\bk,m\bq}(\bR+\bT)
=\ee^{\ii(\bq-\bk)\cdot\bT}\,
\varrho_{n\bk,m\bq}(\bR).
\label{eq:neq_transition_density_phase}
\end{equation}
An allowed unequal-\(k\) transition density acquires the phase \(\ee^{\ii(\bq-\bk)\cdot\bT}\) from cell to cell.
For successive translations separated by \(\Delta\bT\) along the allowed lattice direction, the phase increment is
\(\Delta\phi=(\bq-\bk)\cdot\Delta\bT\). Its zeroth translational moment cancels because \(S_T=0\), whereas the
position-weighted first translational moment remains finite along that
direction. The unequal-\(k\) block therefore represents the first moment
of a cell-to-cell phase progression. It is the finite-domain counterpart
of momentum-space transport, not a separate optical-transition channel.

\section{Consistency of the finite-BvK representation}
\label{sec:consistency}

For a fixed mesh, active-band subspace, and coordinate branch,
Eq.~\eqref{eq:finite_matrix_definition} is the Galerkin representation of
coordinate multiplication. The matrix itself is branch and band-frame
dependent. Observables remain consistent only when the position matrix,
Hamiltonian, density matrix, and relaxation operator are transformed
together.

\subsection{Coordinate branch and equal-\(k\) structure}

The continuous coordinate branch \(0\le h_\mu+s_\mu<N_\mu\), with \(\mu=1,2,3\), has geometric center
\begin{equation}
\mathbf R_{c,\mathbf N}=\frac12\sum_{\mu=1}^{3}N_\mu\ba_\mu.
\end{equation}
Matrices obtained with different branches must be referred to the same physical origin. The branch-aligned representation is
\begin{equation}
\boldsymbol{\mathcal R}^{\mathrm{cs}}_{\mathbf N}
=\boldsymbol{\mathcal R}_{\mathbf N}
-\mathbf R_{c,\mathbf N}\mathbf 1,
\end{equation}
followed by
\begin{equation}
\widetilde{\boldsymbol{\mathcal R}}_{\mathbf N}
=\mathcal U_{\mathbf N}^{\dagger}
\boldsymbol{\mathcal R}^{\mathrm{cs}}_{\mathbf N}
\mathcal U_{\mathbf N}.
\label{eq:finite_operator_action}
\end{equation}
where \(\mathcal U_{\mathbf N}\) is the basis transformation induced by the origin shift and branch cut. 
For a rigid shift of the coordinate origin by \(\mathbf R_{c,\mathbf N}\),
\begin{equation}
[\mathcal U_{\mathbf N}]_{n\bk,m\bq}
=\delta_{nm}\delta_{\bk\bq}
\exp[-\ii\bk\cdot\mathbf R_{c,\mathbf N}].
\end{equation}
Subtracting this geometric branch center without the accompanying basis transformation removes the extensive diagonal 
constant but leaves the branch cut inconsistent. The BvK normalization and 
the uniform \(k\)-mesh weight \(1/N_R\) enter every matrix contraction.

Define the dimensionless intracell matrix
\begin{equation}
X^d_{nm}(\bk)=\langle u_{n\bk}|s_d|u_{m\bk}\rangle_{\cell}.
\label{eq:intracell_X}
\end{equation}
All-electron orthonormality gives the raw equal-\(k\) block
\begin{equation}
\boldsymbol{\mathcal R}_{\mathbf N;nm}(\bk,\bk)
=
\sum_d\ba_d
\left[
X^d_{nm}(\bk)+\frac{N_d-1}{2}\delta_{nm}
\right],
\label{eq:equal_k_compact}
\end{equation}
and the center-subtracted equal-\(k\) block
\begin{equation}
\boldsymbol{\mathcal R}^{\mathrm{cs}}_{\mathbf N;nm}(\bk,\bk)
=
\sum_d\ba_d
\left[
X^d_{nm}(\bk)-\frac12\delta_{nm}
\right].
\label{eq:center_subtracted_equal_k}
\end{equation}
Equation~\eqref{eq:center_subtracted_equal_k} contains only the subtraction in
 Eq.~\eqref{eq:finite_operator_action}; the fully branch-aligned representation 
 additionally includes the basis transformation \(\mathcal U_{\mathbf N}\). 
 Thus the extensive branch term occurs only in the intraband diagonal; for \(n\ne m\), the block is the
intracell coherence-displacement kernel of Eq.~\eqref{eq:coherence_displacement_kernel}.

\subsection{Near-diagonal scaling}

For an allowed pair \(\bq=\bk+\delta\bb_d\), define the all-electron overlap
\begin{equation}
S_{nm}^{(d)}(\delta)
=
\left\langle u_{n\bk}\middle|
\ee^{\ii2\pi\delta s_d}
\middle|u_{m,\bk+\delta\bb_d}\right\rangle_{\cell}.
\end{equation}
In a differentiable frame,
\begin{align}
S_{nm}^{(d)}(\delta)
&=\delta_{nm}+\delta\,\Lambda^d_{nm}(\bk)+O(\delta^2), \nonumber\\
\Lambda^d_{nm}(\bk)
&=\left.\frac{\partial S_{nm}^{(d)}}{\partial\delta}\right|_{0}.
\label{eq:overlap_slope}
\end{align}
Using
\begin{equation}
-\frac{1}{1-\ee^{\ii2\pi\delta}}
=
\frac{1}{\ii2\pi\delta}-\frac12+O(\delta),
\label{eq:phase_factor_small_delta}
\end{equation}
yields
\begin{equation}
\lim_{\delta\to0}
\boldsymbol{\mathcal R}_{nm}(\bk,\bk+\delta\bb_d)
=
\frac{\ba_d}{\ii2\pi}\Lambda^d_{nm}(\bk),
\qquad n\ne m,
\label{eq:neq_interband_limit}
\end{equation}
and
\begin{equation}
\boldsymbol{\mathcal R}_{nn}(\bk,\bk+\delta\bb_d)
=
\ba_d\left[
\frac{1}{\ii2\pi\delta}
+\frac{\Lambda^d_{nn}(\bk)}{\ii2\pi}
-\frac12+O(\delta)
\right].
\label{eq:neq_intraband_limit}
\end{equation}
Thus the intraband block scales as \(N_d\) for a nearest-neighbor mesh
 separation \(\delta\sim1/N_d\), whereas orthogonality leaves a finite interband limit. 
 The limit \(\delta\to0\) is understood along a sequence of increasingly dense BvK-compatible meshes.

The relation to Berry-connection formulations is 
seen by differentiating the overlap in Eq.~\eqref{eq:overlap_slope}.
We define the non-Abelian connection in the reduced coordinate \(\kappa_d\) by
\begin{equation}
\mathcal A^d_{nm}(\bk)
=
\ii\left\langle u_{n\bk}\middle|\frac{\partial}{\partial\kappa_d}u_{m\bk}\right\rangle_{\cell}.
\label{eq:berry_connection_definition}
\end{equation}
Then
\begin{equation}
\Lambda^d_{nm}(\bk)
=
\ii\left[2\pi X^d_{nm}(\bk)-\mathcal A^d_{nm}(\bk)\right].
\label{eq:lambda_berry_relation}
\end{equation}
The explicit intracell-coordinate term reflects the chosen finite branch,
whereas \(\mathcal A^d\) is the crystal-momentum-derivative contribution.
In the local limit, the finite-\(Q\) theory recovers the familiar
derivative--Berry description. At finite momentum separation, it retains
additional amplitude, phase, and PAW information
\cite{blount1962,si2025position}.

\subsection{Limiting cases and Hermiticity}

For a single plane-wave band on \(0\le x<L\), direct integration gives
\begin{equation}
\langle k_{\ell}|x|k_{\ell'}\rangle_{[0,L)}
=
\begin{cases}
L/2, & \ell=\ell',\\[4pt]
1/[\ii(k_{\ell'}-k_{\ell})], & \ell\ne\ell'.
\end{cases}
\label{eq:free_electron_finite_matrix}
\end{equation}
On the centered branch \(-L/2\le x<L/2\),
\begin{equation}
\langle k_{\ell}|x|k_{\ell'}\rangle_{[-L/2,L/2)}
=
\begin{cases}
0, & \ell=\ell',\\[4pt]
\dfrac{(-1)^{\ell'-\ell}}{\ii(k_{\ell'}-k_{\ell})}, & \ell\ne\ell',
\end{cases}
\label{eq:free_electron_centered_matrix}
\end{equation}
where \((-1)^{\ell'-\ell}\) is the origin-shift phase generated by Eq.~\eqref{eq:finite_operator_action}. 
The finite-BvK matrix reproduces both identities, contains the predicted
number of allowed \(k\)-pairs, and has vanishing forbidden blocks.

The complete all-electron matrix also satisfies
\begin{equation}
\mathcal R^\alpha_{nm}(\bk,\bq)
=\left[\mathcal R^\alpha_{mn}(\bq,\bk)\right]^*,
\label{eq:hermiticity_check}
\end{equation}
This relation simultaneously constrains the
pseudo and PAW terms, their atomic phases, and the reversed-\(k\)
convention.

\subsection{Band-frame covariance and alignment}

A unitary rotation \(W(\bk)\) within the retained subspace transforms each Cartesian component as
\begin{equation}
\mathcal R^\alpha(\bk,\bq)\rightarrow
W^\dagger(\bk)\mathcal R^\alpha(\bk,\bq)W(\bq),
\label{eq:position_gauge_covariance}
\end{equation}
and the density matrix as
\begin{equation}
\rho(\bq,\bk)\rightarrow
W^\dagger(\bq)\rho(\bq,\bk)W(\bk).
\end{equation}
Hence \(\operatorname{Tr}[\rho(\bq,\bk)\mathcal R^\alpha(\bk,\bq)]\) is invariant, 
although individual blocks remain frame dependent.

Independent diagonalizations assign arbitrary phases and non-Abelian rotations at neighboring \(\bk\) points.
 Let \(N_b\) states span the retained subspace and write
\begin{equation}
|\chi_{a\bk}\rangle
=
\sum_{n=1}^{N_b}|\psi_{n\bk}\rangle G_{na}(\bk).
\end{equation}
Along an ordered mesh path, form the cell-periodic all-electron PAW overlap
\begin{equation}
O_{nm}(\bk_{p-1},\bk_p)
=
\left\langle u_{n\bk_{p-1}}\middle|u_{m\bk_p}\right\rangle_{\cell}^{\mathrm{AE,PAW}}.
\label{eq:neighbor_overlap}
\end{equation}
The cell-periodic overlap is required here because the inner product of complete Bloch states over the BvK domain vanishes for distinct mesh \(k\) points.
If the frame at \(\bk_{p-1}\) is fixed, define \(M=G^\dagger(\bk_{p-1})O\) and compute \(M=U_{\mathrm{s}}\Sigma V_{\mathrm{s}}^\dagger\). The parallel-transport choice
\begin{equation}
G(\bk_p)=V_{\mathrm{s}}U_{\mathrm{s}}^\dagger,
\label{eq:parallel_transport_gauge}
\end{equation}
maximizes the overlap with the preceding point. Small singular values indicate leakage from the retained manifold. A general operator is then transformed as
\begin{equation}
\mathcal O^g(\bk,\bq)=G^\dagger(\bk)\mathcal O^{\mathrm{raw}}(\bk,\bq)G(\bq),
\label{eq:gauge_transformation}
\end{equation}
including
\begin{equation}
H_0^g(\bk)=G^\dagger(\bk)\,\mathrm{diag}[\epsilon_{1\bk},\ldots,\epsilon_{N_b\bk}]\,G(\bk).
\label{eq:gauge_transformed_h0}
\end{equation}
On a multidimensional mesh, loop holonomy encodes the non-Abelian geometry
of the transported manifold \cite{marzari1997}.

\section{Length-gauge dynamics}

The propagation uses the branch-aligned, PAW-reconstructed
eight-band position matrix of Eq.~\eqref{eq:finite_operator_action}.
The branch transformation \(\mathcal U_{\mathbf N}\) is applied first, followed by the non-Abelian frame alignment \(G(\bk)\). 
Projection onto the laser polarization therefore gives
\begin{equation}
D^g_{ab}(\bk,\bq)=\mathbf e_{\mathrm{pol}}\cdot
\left[G^\dagger(\bk)\widetilde{\boldsymbol{\mathcal R}}_{\mathbf N}(\bk,\bq)G(\bq)\right]_{ab}.
\label{eq:aligned_dipole_matrix}
\end{equation}
The field-coupled Hamiltonian is
\begin{equation}
H^g_{a\bk,b\bq}(t)=
\delta_{\bk\bq}\left[H_0^g(\bk)\right]_{ab}
+eE(t)D^g_{ab}(\bk,\bq),
\label{eq:full_hamiltonian}
\end{equation}
and the density matrix in the composite band--momentum basis obeys
\begin{equation}
\ii\hbar\dot\rho^g(t)
=[H^g(t),\rho^g(t)]
-\ii\hbar\,\mathcal L_{\mathrm{rel}}^g[
\rho^g(t)-\rho^{g(0)}].
\label{eq:full_density_matrix}
\end{equation}
Here \(\rho^{g(0)}\) is the field-free equilibrium density matrix in the
aligned frame. In the field-free eigenstate frame, with
\(\delta\rho^{\mathrm{eig}}=\rho^{\mathrm{eig}}-\rho^{\mathrm{eig}(0)}\),
the phenomenological relaxation is defined elementwise as
\begin{equation}
\left[\mathcal L_{\mathrm{rel}}^{\mathrm{eig}}
(\delta\rho^{\mathrm{eig}})\right]_{n\bk,m\bq}
=
\begin{cases}
\Gamma_1\,\delta\rho^{\mathrm{eig}}_{n\bk,n\bk},
& n=m\ \text{and}\ \bk=\bq,\\[3pt]
\Gamma_2\,\delta\rho^{\mathrm{eig}}_{n\bk,m\bq},
& \text{otherwise}.
\end{cases}
\label{eq:relaxation_definition}
\end{equation}
Thus all interband and unequal-\(k\) coherences decay with \(\Gamma_2\),
while diagonal population deviations relax with \(\Gamma_1\). To preserve
basis covariance, define the block-diagonal transformation
\([\mathcal G]_{n\bk,a\bq}=\delta_{\bk\bq}G_{na}(\bk)\) and use
\begin{equation}
\mathcal L_{\mathrm{rel}}^g[\delta\rho^g]
=\mathcal G^\dagger\,
\mathcal L_{\mathrm{rel}}^{\mathrm{eig}}
[\mathcal G\delta\rho^g\mathcal G^\dagger]\,
\mathcal G .
\label{eq:relaxation_frame_transformation}
\end{equation}

We consider a Gaussian-envelope electric field
\begin{equation}
E(t)=E_0\exp\left[-\frac12\left(\frac{t-t_0}{\sigma}\right)^2\right]
\sin[\omega_0(t-t_0)],
\label{eq:laser_field}
\end{equation}
with \(\hbar\omega_0=E_{\mathrm{ph}}\), \(t_0=(N_t-1)\Delta t/2\), and \(\sigma=\mathrm{FWHM}/[2\sqrt{2\ln2}]\), 
where FWHM refers to the field-amplitude envelope. We use the field-induced
density matrix \(\delta\rho^g(t)=\rho^g(t)-\rho^{g(0)}\). The corresponding
change in dipole moment per primitive cell, equivalently the induced
macroscopic polarization multiplied by the primitive-cell volume, is
\begin{equation}
\Delta p_{\mathrm{cell}}(t)=-\frac{g_s e}{N_R}
\sum_{\bk,\bq}
\operatorname{Re}\,\operatorname{Tr}\!\left[
\delta\rho^g(\bq,\bk;t)D^g(\bk,\bq)\right].
\label{eq:dipole_full}
\end{equation}
Here \(e>0\) is the elementary charge and \(g_s=2\) is the spin-degeneracy
factor of spin-unpolarized Si.
Equal- and unequal-\(k\) coherences contribute together to the total dipole
response. With $t_j=j\Delta t$ and a Hann window \(w(t_j)\),
\begin{equation}
\Delta p_{\mathrm{cell}}(\omega)=\Delta t\sum_{j=0}^{N_t-1}
w(t_j)\Delta p_{\mathrm{cell}}(t_j)\ee^{-\ii\omega t_j},
\end{equation}
\begin{equation}
w(t_j)=\frac12\left[1-\cos\left(\frac{2\pi j}{N_t-1}\right)\right],
\end{equation}
and the cell current in frequency space is
\begin{equation}
J_{\mathrm{cell}}(\omega)=-\ii\omega\Delta p_{\mathrm{cell}}(\omega).
\label{eq:cell_current}
\end{equation}
The spectra below use the current-equivalent signal
\begin{equation}
\mathcal S_J(\omega)=|J_{\mathrm{cell}}(\omega)|^2
=|\omega\Delta p_{\mathrm{cell}}(\omega)|^2.
\label{eq:hhg_signal}
\end{equation}
This convention differs from a radiated-field intensity, which contains
one additional factor of \(\omega^2\); we retain the current convention
used to generate the displayed spectra.
For the field scans, the spectral weight assigned to harmonic order $N$ is
\begin{equation}
Y_N=\hbar\int_{\Omega_N}\mathcal S_J(\omega)\,\dd\omega.
\label{eq:harmonic_yield}
\end{equation}
The nonoverlapping angular-frequency window is
\begin{equation}
\Omega_N=
\left[(N-\tfrac12)\omega_0,(N+\tfrac12)\omega_0\right],
\label{eq:harmonic_window}
\end{equation}
and the same relative window is used for all field amplitudes at a given
photon energy. The factor \(\hbar\) converts the angular-frequency integral
to an energy integral.

We describe diamond-cubic Si in the PAW formalism
\cite{kresse1999,kresse1996} using the two-atom primitive cell with
lattice vectors \(a_0(0,\tfrac12,\tfrac12)\),
\(a_0(\tfrac12,0,\tfrac12)\), and
\(a_0(\tfrac12,\tfrac12,0)\), and Si atoms at fractional coordinates
\((0,0,0)\) and \((\tfrac14,\tfrac14,\tfrac14)\), with
\(a_0=5.43~\text{\AA}\). A \(300\)-eV plane-wave cutoff and a uniformly
shifted \(32\times32\times32\) full-Brillouin-zone mesh are used throughout.
The propagated manifold contains four valence and four conduction bands.
Within the frozen-core approximation, the position matrix was reconstructed
at the all-electron valence level and transformed to the smooth non-Abelian
frame described above.

The primary drive has \(E_{\mathrm{ph}}=0.35\)~eV, a 100-fs
field-amplitude-envelope FWHM, polarization
\(\mathbf e_{\mathrm{pol}}=(1,1,0)/\sqrt2\), and peak fields of
\(0.04\)--\(0.12~\mathrm{V}/\text{\AA}\). We use
\(\hbar\Gamma_1=0.005\)~eV, \(\hbar\Gamma_2=0.080\)~eV,
\(\Delta t=0.05\)~fs, and \(N_t=16\,384\). The \(1.2\)-eV drive has the
same polarization, pulse envelope, and material parameters.

\section{Results}

\subsection{Finite-momentum control of HHG in Si}

\begin{figure}[t]
    \centering
    \includegraphics[width=\columnwidth]{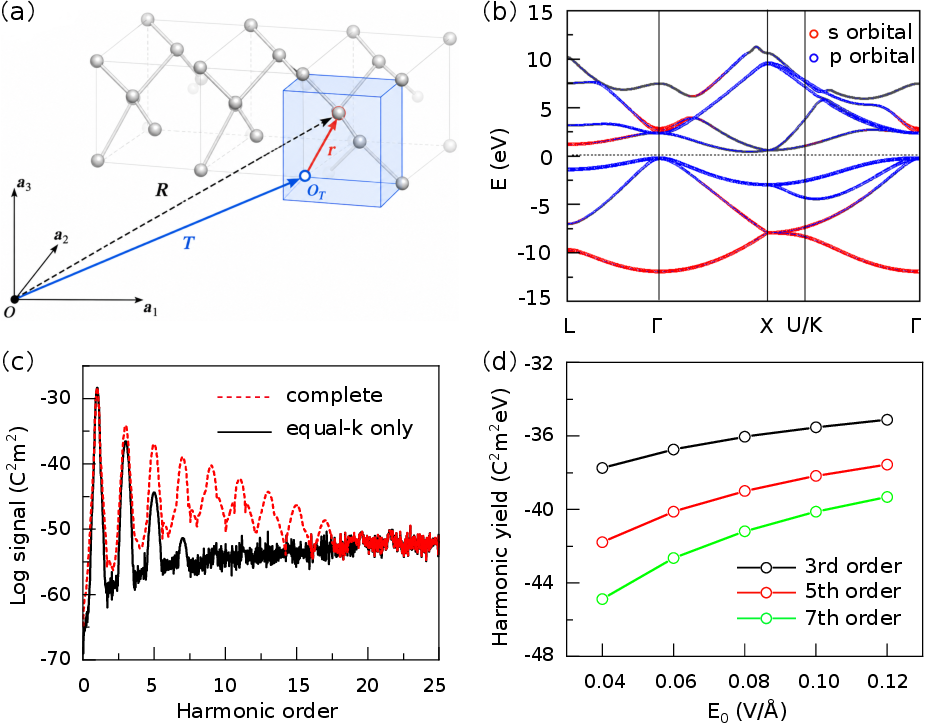}
    \caption{Si electronic structure and low-photon-energy high-harmonic
    response. (a) Diamond-cubic 
    structure and coordinate partition $\bR=\bT+\br$. (b) Orbital-resolved band structure of  Si.
     (c) HHG spectrum at $E_{\mathrm{ph}}=0.35$~eV and $E_0=0.10~\mathrm{V}/\text{\AA}$ 
     for the full two-momentum position matrix (red dashed) and its
     equal-\(k\)-restricted form (black solid). (d) Peak-field dependence
     of the third-, fifth-, and seventh-harmonic yields calculated with the
     full two-momentum position matrix.}
    \label{fig:si}
\end{figure}

Figure~\ref{fig:si} summarizes the Si electronic structure and the
low-photon-energy nonlinear optical response. 
Panel (a) shows the decomposition $\bR=\bT+\br$ for diamond-cubic Si with lattice constant $a_0=5.43~\text{\AA}$~\cite{hom1975}, 
and panel (b) shows the orbital-resolved dispersion of the eight-band
manifold. The indirect gap is about \(0.7\)~eV, and the smallest
direct separation along the plotted high-symmetry path is about \(2.5\)~eV.
These energies define the electronic manifold sampled by the field.

Figure~\ref{fig:si}(c) reveals the central physical result. At
\(E_{\mathrm{ph}}=0.35\)~eV, finite-momentum coherence reorganizes the
fifth and higher harmonics, enhancing some orders and suppressing others
through coherent interference. This information is absent from a purely
local Berry description but enters directly into the measurable nonlinear
response. The field dependence in Fig.~\ref{fig:si}(d) shows how the same
finite-momentum physics grows with the driven electronic motion.

Table~\ref{tab:matrix-elements} reports representative off-diagonal
equal-\(k\) intracell displacement matrix elements at \(L\). Their
Cartesian norms are \(0.192\), \(0.436\), \(0.237\), and
\(0.235~\text{\AA}\) for transitions from band 4 to bands 5--8,
respectively. The componentwise values obey Hermiticity and the PAW
reconstruction. The direct coordinate representation avoids momentum
derivatives and interband energy denominators, while the smooth multiband
frame preserves covariance during the field-driven evolution.

\begin{table*}[t]
    \centering
    \caption{Off-diagonal equal-\(k\) intracell coherence displacements at $L$, in \(\text{\AA}\).}
    \label{tab:matrix-elements}
    \setlength{\tabcolsep}{10pt}
    \setlength{\arrayrulewidth}{0.6pt}
    \setlength{\doublerulesep}{1.2pt}
    \renewcommand{\arraystretch}{1.18}
    \begin{tabular}{c r r r r r r}
        \hline\hline
        & \multicolumn{2}{c}{$x$} & \multicolumn{2}{c}{$y$} & \multicolumn{2}{c}{$z$} \\
        \cline{2-3}\cline{4-5}\cline{6-7}
        Band pair
        & $\operatorname{Re}$ & $\operatorname{Im}$
        & $\operatorname{Re}$ & $\operatorname{Im}$
        & $\operatorname{Re}$ & $\operatorname{Im}$ \\
        \hline
        $4\text{--}5$ & $-0.0236$ & $0.0081$ & $0.0282$ & $0.1267$ & $0.1386$ & $-0.0103$ \\
        $4\text{--}6$ & $-0.0260$ & $0.0021$ & $-0.3676$ & $0.0203$ & $-0.2323$ & $-0.0070$ \\
        $4\text{--}7$ & $-0.0307$ & $0.0588$ & $-0.0716$ & $-0.0168$ & $0.2068$ & $-0.0605$ \\
        $4\text{--}8$ & $0.1345$ & $-0.0169$ & $0.0994$ & $0.0553$ & $-0.0419$ & $0.1494$ \\
        \hline
    \end{tabular}
\end{table*}

We next consider the unequal-\(k\) sector. The finite BvK sums do not
produce a dense all-to-all matrix 
in $\bk$ and $\bq$; instead they impose an exact allowed-pair structure. 
For a cubic $M^3$ mesh, the ordered pair counts are
\begin{align}
N_{\mathrm{all}}&=M^6,\\
N_{\mathrm{eq}}&=M^3,\\
N_{\mathrm{neq}}&=3M^3(M-1),\\
N_{\mathrm{forbidden}}&=M^6-M^3-3M^3(M-1).
\label{eq:neq_pair_counts}
\end{align}
These are ordered counts of structurally allowed vector blocks. A
particular Cartesian component or polarization projection can vanish when
its contraction with the corresponding primitive vector is zero.
For every $\bk$, the allowed unequal-\(k\) partners lie on three 
reciprocal-space lines: exactly one reduced component differs,
 while the two transverse components obey Eq.~\eqref{eq:transverse_compensation}. 
Table~\ref{tab:finite-bvk-selection} shows the momentum-pair multiplicities
implied by Eq.~\eqref{eq:neq_pair_counts}. Their \(M^4\) growth follows from
the line support of the translational first moment, in contrast to the
\(M^6\) growth of all possible pairs. Together with the free-electron,
Hermiticity, and branch-transformation relations, this fixes the structure
of the projected coordinate matrix for the chosen finite mesh, band
subspace, coordinate branch, and PAW representation.

\begin{table}[t]
\centering
\caption{Ordered momentum pairs selected by the finite-BvK position kernel
on cubic \(M^3\) meshes. The calculations reported here use \(M=32\).}
\label{tab:finite-bvk-selection}
\scriptsize
\setlength{\tabcolsep}{2.6pt}
\setlength{\arrayrulewidth}{0.6pt}
\setlength{\doublerulesep}{1.2pt}
\renewcommand{\arraystretch}{1.12}
\begin{tabular}{c r r r r}
\hline\hline
$M^3$ & {$N_{\mathrm{all}}$} & {$N_{\mathrm{eq}}$} & {$N_{\mathrm{neq}}$} & {$N_{\mathrm{forbidden}}$} \\
\hline
$2^3$  & 64        & 8     & 24      & 32        \\
$3^3$  & 729       & 27    & 162     & 540       \\
$12^3$ & 2985984   & 1728  & 57024   & 2927232   \\
$16^3$ & 16777216  & 4096  & 184320  & 16588800  \\
$20^3$ & 64000000  & 8000  & 456000  & 63536000  \\
$24^3$ & 191102976  & 13824 & 953856  & 190135296  \\
$32^3$ & 1073741824 & 32768 & 3047424 & 1070661632 \\
\hline
\end{tabular}
\end{table}

Equation~\eqref{eq:direct_matrix_general} shows that this response originates
from the translational first moment carried by unequal-\(k\) coherence
\cite{blount1962,houston1940,osika2017,parks2020}. Its line-like momentum
structure converts coherence across finite Brillouin-zone separations into
the interference that controls the harmonic spectrum
\cite{vampa2015,golde2008,wu2015,hohenleutner2015,
tancogne2017coupled,navarrete2019,he2021}.

\begin{figure}[t]
    \centering
    \includegraphics[width=\columnwidth]{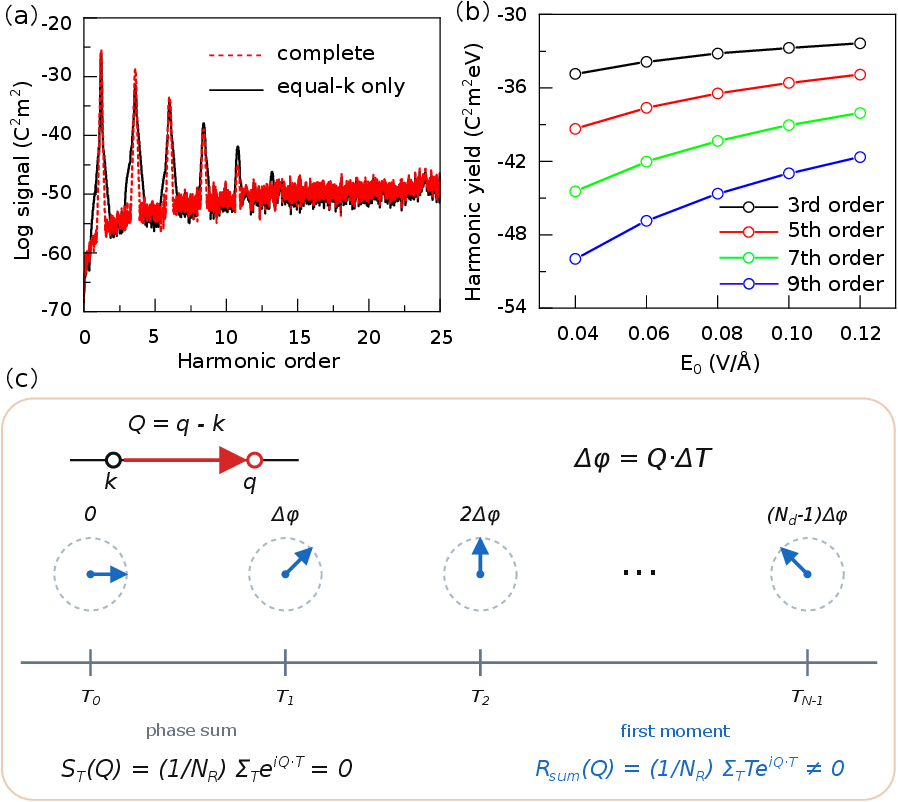}
    \caption{Photon-energy dependence and real-space interpretation. (a)
    HHG spectrum at $E_{\mathrm{ph}}=1.2$~eV 
    and $E_0=0.10~\mathrm{V}/\text{\AA}$ for the full two-momentum
    position matrix (red dashed) and its equal-\(k\)-restricted form
    (black solid). (b) Peak-field dependence of the third-, fifth-, seventh-,
    and ninth-harmonic yields calculated with the full two-momentum position
    matrix. (c) Coherence-dipole mechanism: the phase sum of an allowed
    unequal-\(k\) coherence vanishes, while position weighting leaves a
    finite first moment.}
    \label{fig:frequency-picture}
\end{figure}

\subsection{Field-driven momentum scale}

The comparison with \(E_{\mathrm{ph}}=1.2\)~eV shows how the field-driven
momentum excursion controls the finite-momentum response. The acceleration
theorem gives
\begin{equation}
\Delta\bk(t)=-\frac{e}{\hbar}\int_{-\infty}^{t}
\mathbf E(t')\,\dd t',
\label{eq:houston_trajectory}
\end{equation}
and motivates the excursion-amplitude estimate
\begin{equation}
\Delta k_{\mathrm{amp}}
=\max_t|\Delta\bk(t)|
\simeq\frac{eE_0}{\hbar\omega_0}.
\label{eq:frequency_excursion_comparison}
\end{equation}
At \(E_0=0.10~\mathrm{V}/\text{\AA}\), Eq.~\eqref{eq:frequency_excursion_comparison}
gives \(0.286~\text{\AA}^{-1}\) at \(0.35\)~eV and
\(0.083~\text{\AA}^{-1}\) at \(1.2\)~eV.

At \(1.2\)~eV, the full and equal-\(k\)-restricted spectra nearly
coincide at low harmonic orders and show only moderate differences at
higher orders [Fig.~\ref{fig:frequency-picture}(a)]. The smaller excursion
therefore produces the smaller sensitivity to unequal-\(k\) coherence. The
two spectra identify \(eE_0/(\hbar\omega_0)\) as the momentum scale that
organizes the frequency dependence.

The near-perturbative field scaling in
Fig.~\ref{fig:frequency-picture}(b) further distinguishes this local regime
from the finite-momentum reorganization found at lower frequency.

Equation~\eqref{eq:neq_transition_density_phase} gives a direct real-space
interpretation of the coherence dipole. The transition density in
neighboring primitive cells has the same spatial form but acquires the phase
\(\exp[\ii(\bq-\bk)\cdot\bT]\). Its unweighted sum over the BvK domain
vanishes because \(S_T=0\). Multiplication by the cell translation \(\bT\),
however, leaves a finite first moment along the direction in which \(\bk\)
and \(\bq\) differ, as illustrated in Fig.~\ref{fig:frequency-picture}(c).
Coupled to the driven electronic coherence, this surviving first moment
converts a momentum-space phase texture into macroscopic polarization and
changes the interference that generates the harmonic spectrum.

\subsection{Finite-\(Q\) geometry beyond the local limit}

\begin{figure}[t]
    \centering
    \includegraphics[width=\columnwidth]{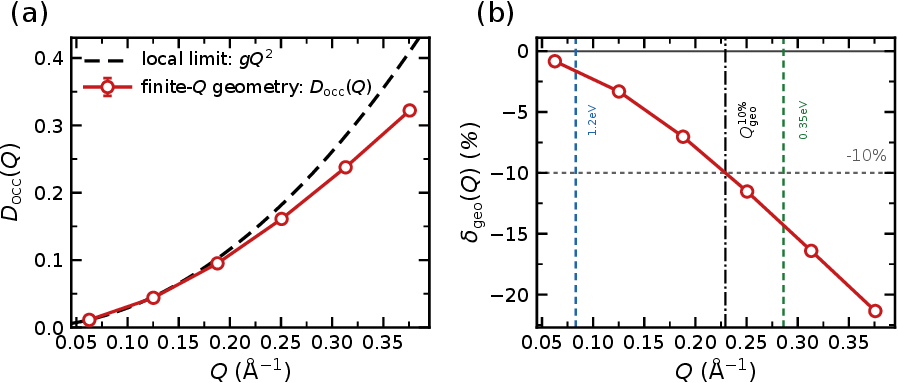}
    \caption{Finite-momentum geometry in Si. (a) Occupied-subspace distance
    \(D_{\mathrm{occ}}(Q)\) and its local quantum-metric form \(gQ^2\).
    (b) Departure from local geometry. The black line marks the
    characteristic momentum
    \(Q_{\mathrm{geo}}^{10\%}=0.229~\text{\AA}^{-1}\); blue and green
    lines mark the momentum excursions of the \(1.2\)- and \(0.35\)-eV
    drives.}
    \label{fig:finite-q-geometry}
\end{figure}

We now place the field-driven response on an intrinsic geometric scale.
For the four occupied bands, let
\(M_{\mathrm{occ}}(\bk,\bQ)\) denote the PAW-reconstructed overlap
between the cell-periodic subspaces at \(\bk\) and \(\bk+\bQ\). Its
gauge-invariant occupied-subspace distance is
\begin{equation}
D_{\mathrm{occ}}(Q)=N_{\mathrm{occ}}-
\left\langle\operatorname{Tr}\!\left[
M_{\mathrm{occ}}^\dagger(\bk,\bQ)
M_{\mathrm{occ}}(\bk,\bQ)
\right]\right\rangle_{\bk,d},
\label{eq:occupied_chordal_distance}
\end{equation}
where \(N_{\mathrm{occ}}=4\) and the average covers the Brillouin zone and
three symmetry-equivalent primitive reciprocal-space directions. Near
\(Q=0\), the finite-\(Q\) overlap recovers local Berry geometry, while its
distance follows the occupied-subspace quantum metric,
\begin{equation}
D_{\mathrm{occ}}(Q)=gQ^2+O(Q^4),\qquad
\delta_{\mathrm{geo}}(Q)=
\frac{D_{\mathrm{occ}}(Q)-gQ^2}{gQ^2}.
\label{eq:finite_q_geometry}
\end{equation}
Thus \(D_{\mathrm{occ}}\) provides a gauge-invariant measure of the
finite-separation geometry carried by the two-momentum kernel
\cite{marzari1997}.
The \(gQ^2\) term describes the local tangent plane of the occupied-state
manifold; its higher-order departure measures how that manifold bends away
from local Berry geometry at finite momentum.

Figure~\ref{fig:finite-q-geometry}(a) shows that the electronic geometry
recovers the local form near \(Q=0\) and then departs systematically as the
momentum separation grows. We identify
\(Q_{\mathrm{geo}}^{10\%}=0.229~\text{\AA}^{-1}\) as the finite-momentum
crossover where \(|\delta_{\mathrm{geo}}|=10\%\). Its inverse,
\(4.36~\text{\AA}\), is comparable to the Si lattice constant and connects
the momentum-space crossover to an atomic material length.

The two drives fall on opposite sides of this material scale
[Fig.~\ref{fig:finite-q-geometry}(b)]: the \(1.2\)-eV excursion reaches only
\(0.083~\text{\AA}^{-1}\), whereas the \(0.35\)-eV excursion reaches
\(0.286~\text{\AA}^{-1}\). Their relation to the material geometry is
captured by the geometric-excursion ratio
\begin{equation}
\chi_{\mathrm{geo}}
=\frac{\Delta k_{\mathrm{amp}}}{Q_{\mathrm{geo}}^{10\%}}
=\frac{eE_0}{\hbar\omega_0Q_{\mathrm{geo}}^{10\%}}.
\label{eq:geometry_excursion_ratio}
\end{equation}
It is \(0.36\) for the \(1.2\)-eV drive and \(1.25\) for the \(0.35\)-eV
drive. Finite-momentum geometry therefore becomes dynamically active when
the field-driven excursion crosses the material's geometric scale.

This ordering cannot be explained by emitted photon energy alone. Under the
\(0.35\)-eV drive, the strong spectral reorganization begins at the fifth
harmonic, \(1.75\)~eV, below the direct gap along the plotted path. Under the
\(1.2\)-eV drive, the third harmonic already reaches \(3.6\)~eV, above that
gap, yet remains nearly unchanged. The controlling variable is thus the
drive-to-geometry ratio \(\chi_{\mathrm{geo}}\), rather than a simple
band-gap or output-energy condition.

\FloatBarrier

\section{Discussion}

The unequal-\(k\) matrix elements are the finite-momentum content of
electronic position. They transport Bloch amplitudes across the Brillouin
zone and retain the phase relations sampled by a strong field. In the local
limit this structure reduces to the familiar derivative--Berry description;
at finite separation it carries the additional geometry that reorganizes
nonlinear emission
\cite{souza2004,reislohner2022,blount1962,yue2020,houston1940}.

The same physics defines a finite-momentum coherence dipole. Unequal-\(k\)
coherence carries a cell-to-cell phase whose ordinary sum cancels, whereas
position weighting leaves a finite translational moment. Electronic
position therefore converts a density-dark momentum-space phase texture
into a polarization-active degree of freedom. This coherence dipole changes
the interference of momentum-resolved amplitudes, enhancing some harmonics
and suppressing others
\cite{hohenleutner2015,navarrete2019,he2021,yue2020,foldi2017}.

Equation~\eqref{eq:geometry_excursion_ratio} converts this mechanism into a
material-specific strong-field principle. Unlike a momentum excursion
normalized only by the lattice period, \(\chi_{\mathrm{geo}}\) is set by the
electronic geometry of the material itself. Once \(\chi_{\mathrm{geo}}\)
crosses unity, the field resolves finite-distance geometry and activates
the coherence dipole in nonlinear emission. The order-selective harmonic
reorganization shows that HHG acts as a nonlinear amplifier of this hidden
finite-momentum structure
\cite{vampa2015,vampa2014,freeman2022,wu2015,tancogne2017coupled}.

The two-momentum construction extends crystalline response beyond local
Berry geometry. It provides the first material-specific, first-principles
realization of finite-momentum electronic position, derived from the same
PAW wave functions that determine the dynamics and observable spectrum.
HHG is its first spectroscopic demonstration; the underlying theory
establishes a general basis for electronic phenomena governed by coherence
across finite momentum separations.

\FloatBarrier
\section{Conclusion}

We have extended electronic position, and therefore quantum geometry, from
the local Berry description to finite momentum within a first-principles
framework. The resulting unequal-momentum coherence is dark to the zeroth
spatial moment but bright to polarization, forming a coherence dipole that
is absent from local Berry geometry. In Si, the occupied-subspace geometry
defines the material scale
\(Q_{\mathrm{geo}}^{10\%}=0.229~\text{\AA}^{-1}\), and the ratio
\(\chi_{\mathrm{geo}}=\Delta k_{\mathrm{amp}}/Q_{\mathrm{geo}}^{10\%}\)
determines when this finite-momentum physics becomes active. Crossing this
scale strongly reorganizes the fifth and higher harmonics, establishing a
direct connection between hidden finite-momentum coherence and observable
nonlinear response. Because the electronic structure,
finite-\(Q\) position, and dynamics follow from the same PAW wave functions,
the theory connects nonlocal quantum geometry directly to experiment without
effective band models or fitted transition elements. By making
finite-momentum electronic geometry calculable and experimentally
accessible, this work establishes a new foundation for understanding
nonlocal quantum dynamics in crystalline matter.

\end{document}